\documentclass[prl,twocolumn,superscriptaddress,showpacs]{revtex4}

\makeatletter

\begin{document}
\title{Analog of Astrophysical Magnetorotational Instability in a Couette-Taylor Flow of Polymer Fluids}
\author{Stanislav Boldyrev} 
\author{Don Huynh}
\affiliation{Department of Physics, University of Wisconsin-Madison, 1150 University Ave.,  
Madison, WI 53706}
\author{Vladimir Pariev\footnote{Currently at Korea Astronomy and Space Science Institute, 61-1, Hwaam-dong, Yuseong-gu, Daejeon, 305-348, South Korea}}
\affiliation{Department of Physics, University of Wisconsin-Madison, 1150 University Ave.,  
Madison, WI 53706}
\date{June 24, 2008}

\input psfig.sty

\begin{abstract}
We report experimental observation of an instability in 
a Couette-Taylor flow of a polymer fluid in a thin gap between two coaxially rotating cylinders in 
a regime where their angular velocity decreases with the radius while the specific 
angular momentum increases with the radius. In the considered regime, neither the inertial Rayleigh instability nor 
the purely elastic instability are possible. We propose that the observed ``elasto-rotational'' instability is an analog 
of the magnetorotational instability which plays a fundamental role in astrophysical Keplerian accretion disks.  
\pacs{95.30.Qd, 52.30.Cv}
\end{abstract}

\maketitle

{\em Introduction.}---Accretion is a fundamental process in astrophysics by which protostellar objects and stars are formed. Due to gravity, the interstellar gas collapses into thin disks differentially 
rotating around accreting bodies. In order for the gas to further fall onto the central object, angular momentum has to be transported out of the system. Since the molecular viscosity in disks is very small, 
the laminar Keplerian disk cannot loose its angular momentum on astrophysically reasonable timescales. 
The need for much larger, possibly turbulent angular momentum transport was 
identified in~\citet{lynden-bell,shakura-sunyaev}, although it remained unclear what could make a 
hydrodynamically stable Keplerian flow turbulent.     
In the 1990's it was realized that a weak magnetic field existing in accretion disks leads to a quickly growing instability rendering the disks turbulent. This is the  
magneto-rotational instability (MRI) originally 
derived by~\citet{velikhov} and~\citet{chandrasekhar} and rediscovered in astrophysical context 
by~\citet{balbus-hawley}. 

      During the last decade, considerable progress has been made in understanding the effects of such 
instability on differentially rotating flows. A sizable amount of analytic work was devoted to 
linear analysis of instability thresholds~\citep[e.g.,][]{goodman-ji,hughes-tobias,hollerbach-rudiger}.  
Extensive numerical simulations of the nonlinear stage of the magnetorotational instability 
have also been performed, e.g.,~\citep{hawley-gammie-balbus,balbus-hawley2,miller-stone,hawley1,hawley2}, 
however, it still remains a challenge to address the ranges of scales 
relevant for real astrophysical systems. Laboratory investigations of the magneto-rotational 
instability in liquid-metal experiments have been 
proposed~\citep[e.g.,][]{rudiger2,kageyama,willis,noguchi,velikhov_etal}  
and conducted~\citep{sisan,lathrop}. However, large resistivity 
of liquid metals complicates unambiguous laboratory study of 
the magnetorotational instability. Recently, however, the instability was    
observed in a Couette-Taylor liquid metal experiment where 
the helical rather than ``standard'' axial magnetic field was applied by 
external coils~\citep{stefani}. The relevance of this setting for Keplerian 
accretion disks is discussed in~\citep{liu}.

In the present paper we report a laboratory observation of an analog of astrophysical magnetorotational 
instability in an experiment 
using visco-elastic solutions of high molecular weight polymers.  
In a certain range of parameters, the dynamic equations describing visco-elastic polymer fluids 
are identical to the magnetohydrodynamic  
equations describing conducting fluids or plasmas. 
This opens a way to investigate the fundamental astrophysical instability in a simple laboratory setting. 

To explain the physics of the instability, consider two fluid elements rotating 
at different orbits and connected by an elastic string 
(a magnetic field in accretion disks, a polymer in our experiment). The inner element rotates faster, 
therefore it is pulled back by the string. As a result, it loses its angular momentum and falls 
 closer to the center. The outer fluid element is pulled forward, gains angular momentum, 
and goes to a larger orbit. The fluid elements thus move apart stretching the string even more, leading 
to the instability.

Our interest to the problem was motivated by analytic work of \citet{ogilvie-proctor} elucidating 
the analogy between instabilities in Couette-Taylor flows of magnetic and polymer fluids. 
[When our experiment was in progress, we become aware 
of the new paper by \citet{ogilvie-potter} where this analogy is developed in more detail.] 
The experiment provides an intriguing possibility to investigate the 
regime of ``elasto-rotational'' instability and resulting ``elasto-rotational'' 
turbulence in non-Newtonian polymer fluids; such regimes have not been experimentally 
studied before. \\

{\em Magnetohydrodynamics and polymer fluid dynamics.}---
The dynamics of a conducting fluid is described by the set of magnetohydrodynamic  
(MHD) equations: 
\begin{eqnarray}
{}&\partial_t{\bf v}+({\bf v}\cdot \nabla){\bf v}=-\nabla p 
+({\bf B}\cdot \nabla) {\bf B} + \eta \Delta {\bf v} +{\bf F}, 
\label{mhd1} \\
{}&\partial_t {\bf B}+({\bf v}\cdot \nabla){\bf B}-({\bf B}\cdot \nabla){\bf v}=\eta_M \Delta {\bf B},
\label{mhd2}
\end{eqnarray} 
where ${\bf v}({\bf x}, t)$ is the velocity field, ${\bf B}({\bf x}, t)$ is the magnetic field normalized 
by $\sqrt{4\pi \rho}$, 
$p$ is pressure which includes the magnetic part, $\eta$ is fluid viscosity, and $\eta_M$ resistivity. 
We assume that the fluid is incompressible and 
the density is constant, say $\rho=1$.  The external force ${\bf F}({\bf x}, t)$ 
represents mechanisms driving the flow. We also assume 
cylindrical geometry with the coordinates $r$, $\theta$, and $z$,  
where the steady state is described by the azimuthal velocity field~$v_{\theta}(r)$. 
For the gravitational force, $F_r\propto -1/r^2$, the velocity field has 
the Keplerian profile~$v_{\theta}(r)\propto r^{-1/2}$.

As follows from Eq.~(\ref{mhd1}), the back reaction of the magnetic field on 
the flow is described by the Maxwell electromagnetic stress 
tensor~$T_M^{ij}=B^iB^j$.  The evolution equation for this tensor is derived from Eq.~(\ref{mhd2}), 
where we neglect small resistivity~$\eta_M$:
\begin{eqnarray}
\partial_t T_M^{ij}+({\bf v}\cdot \nabla) T_M^{ij}-T_M^{lj}\nabla_lv^i-T_M^{il}\nabla_lv^j=0.
\label{tensor_m}
\end{eqnarray}
The tensor~$T_M^{ij}=B^iB^j$ obeying this equation is said to be ``frozen'' into the flow. 
In the case of polymer fluids, the polymer stress tensor~$T^{ij}_P$ frozen into the flow 
should obey the same equation. 

In contrast with a magnetic fluid, there is no exact equation describing the dynamics of polymer solutions. 
However, in the case when a solution is dilute, one can formulate the constitutive equations based 
on general principles of fluid dynamics~\citep{oldroyd,bird1}.  The stress tensor 
can be represented as a linear sum of the viscous stress of the solvent 
$T^{ij}=\eta (\nabla_i v^j+\nabla_j v^i)$ and the stress $T^{ij}_P$ contributed by the polymer. The polymer 
contribution should generally obey the equation: 
\begin{eqnarray}
T^{ij}_P+\tau D_t T^{ij}_P=\eta_P (\nabla_i v^j+\nabla_j v^i). 
\label{pe}
\end{eqnarray}
Here $\tau$ is the relaxation time of the fluid element, which is related to polymer elasticity, 
and $D_t$ denotes the convective derivative, as in Eq.~(\ref{tensor_m}), 
\begin{eqnarray}  
D_t T_P^{ij}\equiv \partial_t T_P^{ij}+({\bf v}\cdot \nabla) T_P^{ij}-T_P^{lj}\nabla_lv^i-T_P^{il}\nabla_lv^j. 
\label{tensor_p}
\end{eqnarray}

If the relaxation time $\tau$  
is very large compared to a characteristic time of the flow, the second term in the left hand side 
of Eq.~(\ref{pe}) dominates, and the stress is advected by the fluid. In the other 
limit, $\tau \to 0$, 
the polymer is not frozen into the fluid -- it rapidly relaxes to its non-stretched equilibrium configuration 
and  contributes to fluid viscosity, $T^{ij}_P =\eta_P (\nabla_i v^j+\nabla_j v^i)$.  
The dynamics of the velocity field in Eq.~(\ref{tensor_p}) is given by the standard Navier-Stokes equation of motion:
\begin{eqnarray}
\partial_t{\bf v}+({\bf v}\cdot \nabla){\bf v}=-\nabla p + \nabla\cdot {\bf T}_P + \eta \Delta {\bf v} +{\bf F}.   
\label{ns}
\end{eqnarray}
The system~(\ref{pe}), (\ref{tensor_p}) and (\ref{ns}) presents the so-called B model of~\citet{oldroyd}, a constitutive system 
for dilute polymer solutions.  \\

{\em The Ogilvie-Proctor model.}---The analogy of MHD and  polymer fluid instabilities 
in the Couette-Taylor regime was elucidated 
by \citet{ogilvie-proctor}. Following their work, we change the variable: 
 $ 
{T}^{ij}_P \to {\tilde T}^{ij}_P=T^{ij}_P+\frac{\eta_P}{\tau}\delta^{ij},
$ 
where $\delta^{ij}$ is the Kronecker delta. The momentum  
equations (\ref{mhd1}), (\ref{ns})  for magnetic and polymer fluids now have identical forms: 
\begin{eqnarray}
{}&\partial_t{\bf v}+({\bf v}\cdot \nabla){\bf v}=-\nabla p_{M}  
+\nabla \cdot {\bf T}_{M}  + \eta \Delta {\bf v} +{\bf F}, \\
{}& \partial_t{\bf v}+({\bf v}\cdot \nabla){\bf v}=-\nabla p_{P}  
+\nabla \cdot {\tilde {\bf T}}_{P}  + \eta \Delta {\bf v} +{\bf F}, 
\label{momentum-eq}
\end{eqnarray}
where the pressure terms ensure incompressibility of the flows.  
The dynamic equations for the stress tensors are:
\begin{eqnarray}
\partial_t T_M^{ij}+({\bf v}\cdot \nabla) T_M^{ij}-T_M^{lj}\nabla_lv^i-T_M^{il}\nabla_lv^j=\nonumber \\
=\eta_M [B^i\nabla^2 B^j+(\nabla^2B^i) B^j], \label{t_m} \\
\partial_t {\tilde T}_P^{ij}+({\bf v}\cdot \nabla) {\tilde T}_P^{ij}-{\tilde T}_P^{lj}\nabla_lv^i-{\tilde T}_P^{il}\nabla_lv^j=\nonumber \\
=-\frac{1}{\tau}[{\tilde T}^{ij}_P-\frac{\eta_P}{\tau}\delta^{ij}].\label{t_p}
\end{eqnarray}
These equations are identical except for the dissipation terms -- the 
magnetic field diffuses while the polymer 
stress relaxes. 
However, if the magnetic Reynolds number ${\rm Rm}\sim \Omega d^2/\eta_M$ and the Weissenberg 
number ${\rm Wi}= \tau |\partial \Omega/\partial(\ln r)| \sim \tau \Omega$ are large ($\Omega $ being the angular velocity and $d$ 
the gap between cylinders), one can neglect the dissipation terms.  

Denote $R_1$ and $R_2$ as the inner and outer radii, respectively. When the gap is 
narrow, $d/R\ll 1$, the basic velocity flow can be represented 
as $v^{\theta}(r)=v^{\theta}(R_1)+ (r-R_1)\Omega $. The corresponding stationary 
solution of Eq.~(\ref{t_p}) in coordinates $(r, \theta, z)$ then  
has the form \citep{ogilvie-proctor}:
\begin{eqnarray}
{\tilde {\bf T}}_P= \frac{\eta_P}{\tau}\left[
\begin{array}{ccc}
1          & -{\rm Wi}     &   0 \\
-{\rm Wi}  & 2{\rm Wi}^2+1 &   0 \\
0          &     0         &   1 
\end{array}\right].
\label{t_basic}
\end{eqnarray}

There is no exact correspondence of the tensor (\ref{t_basic}) to the 
magnetic tensor ${\bf T}_M$, since (\ref{t_basic}) cannot be represented as a product of two vector fields. 
However, one can introduce a set 
of three auxiliary fields, ${\bf B}_1$, ${\bf B}_2$ and ${\bf B}_3$ such that
\begin{eqnarray} 
{\tilde T}_P^{ij}=B_1^iB_1^j+B_2^iB_2^j+B_3^iB_3^j.
\end{eqnarray}
In this representation, ${\bf B}_1$ and ${\bf B}_2$  have radial and azimuthal components, 
while ${\bf B}_3$ is purely axial \citep{ogilvie-proctor}, 
\begin{eqnarray}
{\bf B}_{1,2}=\left(\frac{\eta_P}{2 \tau}\right)^{1/2}\left[ 
\begin{array}{c}
-1\\
{\rm Wi}\pm({\rm Wi}^2+1)^{1/2}\\
0
\end{array}
\right],
\label{b12}
\end{eqnarray}
\begin{eqnarray}
\quad \quad {\bf B}_{3}=\left(\frac{\eta_P}{\tau}\right)^{1/2}\left[ 
\begin{array}{c}
0\\
0\\
1
\end{array}
\right]. 
\end{eqnarray}

A general analysis of the instability 
requires expansion of the nonlinear equations~(\ref{momentum-eq}) and (\ref{t_p}) in small 
deviations from the basic flow. Depending on what deviations are considered, different ``magnetic fields'' 
play dominant roles. If one assumes  
that the perturbations are axisymmetric ($k_{\theta}=0$), and their wavevectors obey  
$k_z\gg k_r$, then  
the azimuthal and radial fields ${\bf B}_{1,2}$ are not relevant for the instability, and the 
dominant role is played by the axial field~${\bf B}_3$, in direct analogy with the corresponding 
magnetorotational instability.  \\

{\em Experiment.}---In our experiment, a polymer fluid fills the gap between two coaxial cylinders 
rotating in the same direction with different angular 
velocities. The gap is narrow, and the cylinders are driven 
by the same motor with two different gears to approximate the Keplerian velocity 
profile $\Omega(r)\propto r^{-3/2}$.  (In fact, any profile where the outer cylinder rotates 
slower than the inner one, but the specific angular momentum of the outer cylinder is larger 
than that of the inner one is suitable for the considered instability.)

The outer diameter of the inner cylinder is $14$'', the inner diameter of 
the outer cylinder is $15.5$'', and the height of the cylinders is~$2$'. The outer cylinder is transparent and 
the flow is visualized by adding a small amount of highly reflecting  
Kalliroscope particles. The angular velocity of rotation 
can reach  $40\,{\rm rad/s}$, which, for the Keplerian velocity profile translates into 
a shearing rate $\dot \gamma \equiv \partial v_{\theta}/\partial r-v_{\theta}/r\approx 60\, {\rm s}^{-1}$.

For the polymer fluid we choose an aqueous solution of high molecular weight Polyethylene Oxide ($M_W\approx 7,000,000$) obtained  
from DOW Chemical. The experiments were conducted at ambient temperature of $20\,{\rm C}$, 
although the temperature was not precisely controlled. First, we checked that in the studied range 
of angular velocities, the 
hydrodynamic flow without polymer additives was stable. We then performed a series of experiments with different 
concentrations of PolyOx. In each experiment we gradually increased the rotation velocity to obtain the instability threshold.

\begin{figure} [htbp!]
\centerline{\psfig{file=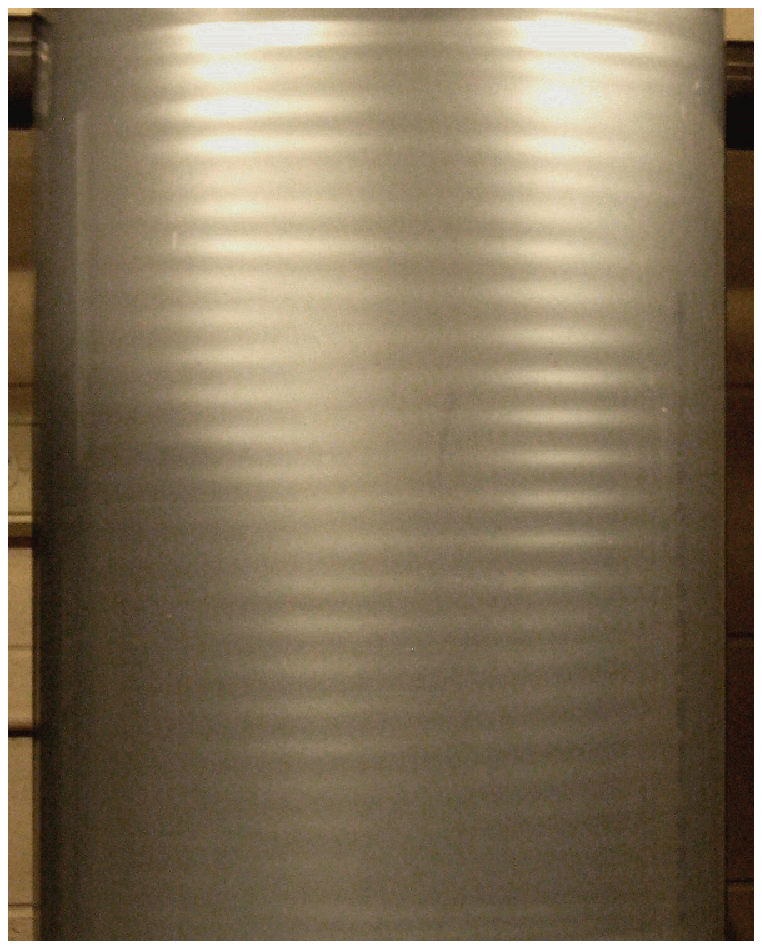,width=1.5in,angle=0}\hskip5mm \psfig{file=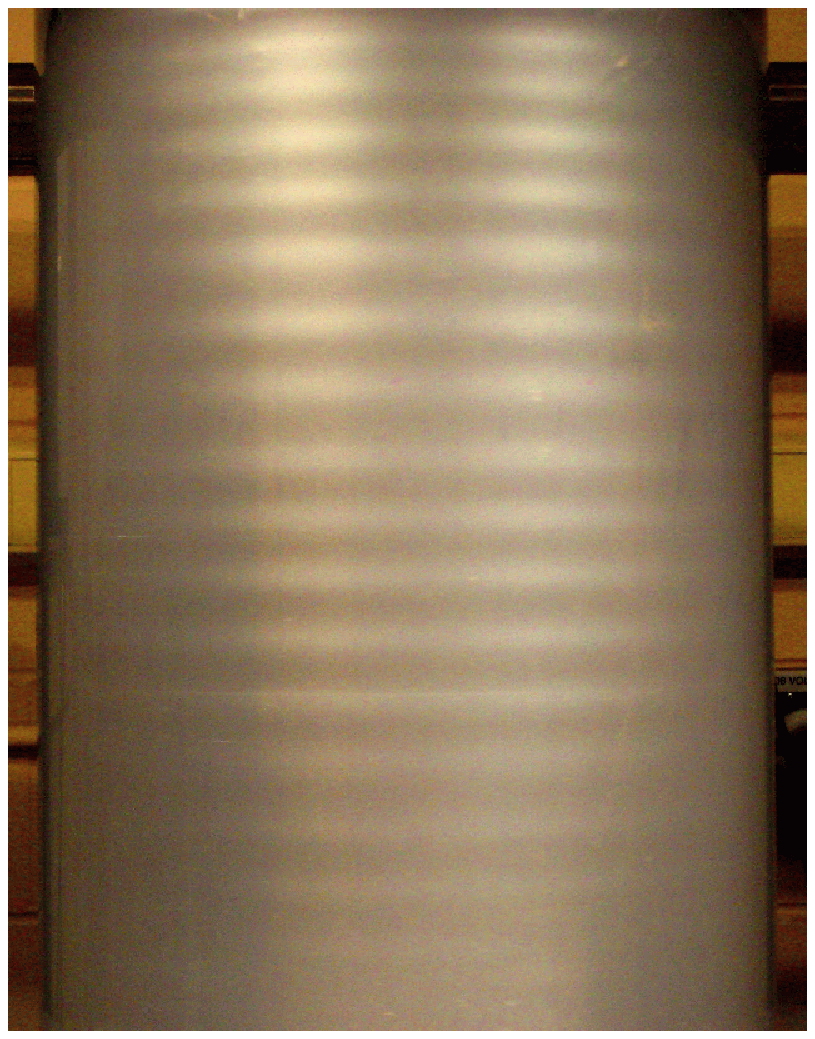,width=1.47in,angle=0}}
\caption{Elasto-rotational instability in visco-elastic Couette-Taylor flow. Left: Most unstable mode is $m=\pm 1$.  Right: Most unstable mode is axisymmetric.}  
\label{fig}
\end{figure}

No instability was observed for concentrations less than about $0.2\%$ by weight, rather,  at very high rotation rates 
turbulence set up at the ends of the cylinders where the Keplerian profile is broken, and propagated over the whole cylinder. 
At higher concentrations, however, the instability did appear.  
At the polymer concentration of $0.25\,\%$, the most unstable mode was a spiral $v(k)\propto \exp(ik_z z + im\theta)$ with azimuthal wave number $m=\pm 1$ and the axial wavelength $\lambda_z\approx 20\,{\rm  mm}$. 
Due to symmetry, the spirals winding up and down are equally probable. In different runs, the flow therefore 
spontaneously broke into regions of $m=1$ and $m=-1$, as e.g. in Fig~(\ref{fig}), left panel.  
The threshold for this instability was 
about $\dot\gamma_c(0.25)\approx 5.6\, {\rm s}^{-1}$.  As the concentration was increased further, the most unstable mode became axisymmetric with the wavelength $\lambda_z\approx 30\,{\rm mm}$.  The threshold for this instability 
was $\dot\gamma_c(0.5)\approx 7.3\, {\rm s}^{-1}$.  The result for $0.5\,\%$ solution is shown in Fig~(\ref{fig}), right panel. In both cases the instability was detectable by eye and the pattern was captured with  a generic digital camera. 

To argue that the observed instability is analogous to the  magnetorotational instability, 
we performed another series of experiments. This time the gears at the cylinders were chosen 
to set up either quasi-Keplerian $\Omega(r)\propto r^{-1.3}$ or ``anti-Keplerian'' 
$\Omega(r)\propto r^{1.3}$ profiles.  This is done to exclude the so-called elastic instability 
that can exist even for small 
Reynolds numbers (${\rm Re}=\Omega d^2/\eta_P$) as long as the Weissenberg 
number exceeds a certain threshold \citep{larson}. The elastic instability 
takes energy from the elastic energy of the flow, and should not essentially depend on the sign 
of $\partial \Omega/\partial r$, while for the MRI the sign of $\partial \Omega/\partial r$ is crucial.

In the ``anti-Keplerian'' case, the only instability that could 
exist is purely elastic instability. For both considered concentrations of PolyOx, we observed 
the instability in the quasi-Keplerian case (analogous to the instability in the Keplerian case), 
however, we  
did not observed any instability in the anti-Keplerian case, even when we increased 
the shearing rates to three times as high as in the Keplerian counterpart. This indicates 
that in our Keplerian case the observed instability is driven by inertia and takes its energy 
from the kinetic energy of the flow. We therefore propose that the observed ``elasto-rotational'' 
instability is analogous to the magnetorotational instability.\\

{\em Discussion.}---
Certain support for our observations is provided by the Ogilvie-Proctor consideration outlined in previous sections. We should be 
cautioned, however, that this model is to some extent phenomenological. It is known that the polymers fluid 
viscosity $\eta_p$ and relaxation time $\tau$ are not constants, but they strongly decrease as the shearing rate increases 
beyond $\dot \gamma \tau \simeq 1$ (the effect of shear-thinning). Besides, a model with single 
relaxation time is often not adequate, and one 
needs to introduce a series of relaxation times describing the relation between the shear rate and stress tensor. More essential, 
however, is the fact that in our case the polymer solution cannot be considered dilute for the used polymer 
concentrations. The application of the theory is therefore limited.

We however found that the theory is in reasonable agreement with the experiment if one substitutes 
the experimentally measured value for unknown viscosity $\eta_p$. Indeed, consider the case of the 
axisymmetric instability observed at polymer concentration $0.5\%$. We measured 
the shear viscosity of the flow $\eta_p$ at the obtained critical shearing rate $\dot \gamma_c$ by measuring 
the viscosity in a Bohlin rheometer using a system of two coaxial cylinders -- 
a scaled down copy of 
the experimental set-up. The inner cylinder was stationary, and the outer cylinder rotated steadily so that the 
shear rate in the gap matched the shear rate in the experiment. The measured viscosity 
was $\eta_p(0.5)\approx 187 {\rm mPa\cdot s}$.  As for the relaxation time, it can be found for very low shearing 
rates using oscillation measurements, giving the value of order $\tau_0(0.5) \sim 1.4\, {\rm s}$. 
The relaxation time is however strongly shear-thinned at the experimental shear rate, so its precise value 
is difficult to evaluate.  

We now substitute the experimental values of the critical shearing rate $\dot \gamma_c$, viscosity $\eta_p$, and the wave 
vector $k_z$, into the linearized Oldroid-B equations. Such linearized equations are derived in 
the limit of small but nonvanishing $d/R$ 
in \citep{larson}; they are bulky and not presented here. We solved these equations numerically. 
The solution confirms that 
the axisymmetric instability with the observed parameters indeed exists if the fluid relaxation time 
is $\tau\sim 0.6 {\rm s}$. This is a reasonable number if shear-thinning is taken into account. 
With this relaxation time we estimate that the instability occurs at ${\rm Wi}=\dot\gamma \tau \sim 4.4$ and ${\rm Re}= \dot \gamma d^2/\eta_P\sim 14$. Moreover, 
when we numerically switched to the anti-Keplerian profile by inverting the sign of the shearing rate $\dot \gamma$, the instability disappeared, which agrees with the experiment.

We note a useful fact that in the axisymmetric case ($m=0$), the instability threshold involves 
only $B_3=\sqrt{\eta_p/\tau}$. In the kinetic theory of dilute polymer solutions the 
ratio $\eta_p/\tau$ is constant and proportional only to polymer 
concentration~\citep[e.g.,][]{doi}. Therefore the ``imposed magnetic field'' ${\bf B}_3$ is  
stable even when both polymer viscosity and relaxation time are shear-thinned by the flow. 
In the non-axisymmetric case ($m\neq 0$), the instability also depends on the azimuthal fields ${\bf B}_{1,2}$.  In principle, it may be possible to design an experiment where such azimuthal field dominates, in even closer analogy with real accretion disks.

In conclusion, based on our results we propose that the analog of magnetorotational instability 
can be experimentally studied in visco-elastic flows of polymer fluids. For a more quantitative analysis  
and for direct comparison with the theory, different polymer solutions whose viscosities are 
not strongly shear-thinned should be used, the so-called Boger fluids~\citep{boger}. This work is in progress.

\acknowledgments
We are grateful to Mark Anderson, Riccardo Bonazza, Cary Forest, Michael Graham, and Daniel Klingenberg for valuable advice and discussions. This work was supported by the Wisconsin Alumni Research 
Foundation. The work of SB and VP was supported by the NSF Center for Magnetic Self-Organization in Laboratory 
and Astrophysical Plasmas at the University of Wisconsin-Madison.


\begin{thebibliography}{99}
\bibitem[Lynden-Bell(1969)]{lynden-bell} Lynden-Bell, D., Nature {\bf 223} (1969) 690.
\bibitem[Shakura \& Syunyaev(1973)]{shakura-sunyaev} Shakura, N. I., Sunyaev, R. A., Astron. Astrophys., {\bf 24} (1973) 337 - 355.
\bibitem[Velikhov(1959)]{velikhov} Velikhov, E. P., Sov. Phys. JETP, {\bf 9} (1959) 995.
\bibitem[Chandrasekhar(1960)]{chandrasekhar} Chandrasekhar, S., Proc. Natl. Acad. Sci. {\bf 46} (1960) 253.
\bibitem[Balbus \& Hawley(1991)]{balbus-hawley} Balbus, S. \& Hawley, J. F., Astrophys. J. {\bf 376} (1991) 214.
\bibitem[Goodman \& Ji(2002)]{goodman-ji} Goodman, J. \& Ji, H., J. Fluid Mech. {\bf 462} (2002) 365.
\bibitem[Hollerbach \& R\"udiger(2005) ]{hollerbach-rudiger} Hollerbach, R. \& R\"udiger, G., Phys. Rev. Lett. {\bf 95} (2005) 124501.
\bibitem[Hughes \& Tobias(2001)]{hughes-tobias} Hughes, D. \& Tobias, S., Proc. R. Soc. Lond. A~{\bf 457} (2001) 1365.
\bibitem[Balbus \& Hawley(1998)]{balbus-hawley2} Balbus, S. \& Hawley, J. F., Rev. Mod. Phys, {\bf 70} (1998) 1-53.
\bibitem[Hawley(2000)]{hawley1} Hawley, J. F., Astrophys. J., {\bf 528} (2000) 462.
\bibitem[Hawley(2001)]{hawley2} Hawley, J. F., Astrophys. J., {\bf 554} (2001) 534. 
\bibitem[Hawley, Gammie \& Balbus(1995)]{hawley-gammie-balbus} Hawley, J. F., Gammie, C. F., \&  Balbus, S. A., 
Astrophys. J. 440, 742 (1995).
\bibitem[Miller \& Stone(2000)]{miller-stone} Miller, K. A., \& Stone, J. M., Astrophys.J., {\bf 534} (2000) 398-419. 
\bibitem[Kageyama, et al.(2004)]{kageyama} Kageyama, A.; Ji, H.; Goodman, J.; Chen, F.; Shoshan, E.,  
J. Phys. Soc. Japan, {\bf 73}, (2004) 2424.
\bibitem[Noguchi, et al.(2002)]{noguchi} Noguchi, K.; Pariev, V. I.; Colgate, S. A.; Beckley, H. F.;\&  Nordhaus, J., Astrophys. J., 
{\bf 575} (2002) 1151-1162.
\bibitem[Velikhov et al(2006)]{velikhov_etal} Velikhov, E. P., Ivanov, A. A., Lakhin, V. P., Serebrennikov, K. S., 
Phys. Lett. A~{\bf 356} (2006) 357.
\bibitem[R\"udiger, Schultz \& Shalybkov(2003)]{rudiger2} R\"udiger, G.; Schultz, M.; Shalybkov, D.,  Phys. Rev. E, {\bf 67} (2003) 046312. 
\bibitem[Willis \& Barenghi(2002)]{willis} Willis, A. P. \& Barenghi, C. F., Astron. Astrophys. {\bf 393} (2002) 339. 
\bibitem[Lathrop(2005)]{lathrop} Lathrop, D., Laboratory sodium experiments modeling astrophysical and geophysical MHD flows, The 47 Annual APS meeting of the Division of Plasma Physics, Oct. 24-28, 2005.
\bibitem[Sisan et al.(2004)]{sisan} Sisan, D.R., Mujica, N., Tillotson, W. A., Huang, Y. M., Dorland, W., Hassam, A. B., Antonsen, T. M., \& Lathrop, D. P., Phys. Rev. Lett.~{\bf 93} (2004) 114502. 
\bibitem[Stefani et al.(2006)]{stefani} Stefani, F., Gundrum, T., Gerbeth, G., R\"udiger, G., Schultz, M., Szklarski, J., \& Hollerbach, R., Phys. Rev. Lett, {\bf 97} (2006) 184502.  
\bibitem[Liu et al.(2006)]{liu} Liu, W; Goodman, J; Herron, I; \& Ji, H., Phys. Rev. E, {\bf 74} (2006) 056302.
\bibitem[Ogilvie \& Proctor(2003)]{ogilvie-proctor} Ogilvie, G. I. \& Proctor, M. R. E.,  J. Fluid Mech. {\bf 476} (2003) 389.  
\bibitem[Ogilvie \& Potter(2008)]{ogilvie-potter} Ogilvie, G. I. \& Potter, A. T., Phys. Rev. Lett~{\bf 100} (2008) 074503. 
\bibitem[Bird, Armstrong \& Hassager(1987)]{bird1} Bird, R. B., Armstrong, R. C., Hassager, O., {\em Dynamics of Polymeric Liquids} 2nd ed. Vol.~1 (Wiley, 1987).
\bibitem[Oldroyd(1950)]{oldroyd} Oldroyd, J. G., Proc. R. Soc. London, A~{\bf 200} (1950) 523.
\bibitem[Larson, Shaqfeh \& Muller(1990)]{larson} Larson, R. G., Shaqfeh, E. S. G., \& Muller S. J.,  J. Fluid Mech. {\bf 218} (1990) 573. 
\bibitem[Doi \& Edwards(1986)]{doi} Doi, M. \& Edwards, S. F., {\em The theory of polymer dynamics} (Oxford University Press, New York, 1986). 
\bibitem[Boger(1978)]{boger} Boger, D. V., J.~Non-Newtonian Fluid Mech., {\bf 3} (1978) 87.















































 
%



 



 










\end{thebibliography}
\end {document}